\documentclass{SciPost}

\hypersetup{
    colorlinks,
    linkcolor={red!50!black},
    citecolor={blue!50!black},
    urlcolor={blue!80!black}
}

\usepackage[bitstream-charter]{mathdesign}
\usepackage{microtype}
\usepackage{xurl}
\usepackage{array}
\DeclareSymbolFont{usualmathcal}{OMS}{cmsy}{m}{n}
\DeclareSymbolFontAlphabet{\mathcal}{usualmathcal}

\fancypagestyle{SPstyle}{
\fancyhf{}
\lhead{\colorbox{scipostblue}{\bf \color{white} ~SciPost Physics Codebases }}
\rhead{{\bf \color{scipostdeepblue} ~Submission }}

\fancyfoot[C]{\textbf{\thepage}}
}

\usepackage{listings}
\usepackage{booktabs}
\usepackage{float} 

\usepackage[notes]{alpseditorial}
\usepackage[normalem]{ulem}

\definecolor{codebg}{RGB}{248,248,245}
\definecolor{codekw}{RGB}{25,60,140}
\definecolor{codestr}{RGB}{30,110,50}
\definecolor{codecomment}{RGB}{120,120,120}
\newcommand{\alps}{\textsc{alps}}
\newcommand{\pyalps}{\texttt{pyALPS}}
\newcommand{\code}[1]{\texttt{#1}}

\begin{document}

\pagestyle{SPstyle}

\begin{center}{\Large \textbf{\color{scipostdeepblue}{
The \alps{} project release 3.0:\texorpdfstring{\\}{ }
open source software for strongly correlated systems
}}}\end{center}

\begin{center}\textbf{
F.~Alet\textsuperscript{1},
T.~Chen\textsuperscript{2},
A.~Feiguin\textsuperscript{3},
E.~Gull\textsuperscript{4,5},
S.~Iskakov\textsuperscript{4},
J.~P.~F.~LeBlanc\textsuperscript{6},
F.~Lin\textsuperscript{7},
A.~Mirmira\textsuperscript{11},
G.~M\"oller\textsuperscript{8},
L.~Pollet\textsuperscript{9},
M.~Rosales\textsuperscript{7},
V.~W.~Scarola\textsuperscript{7$\star$},
H.~Shinaoka\textsuperscript{10},
H.~Terletska\textsuperscript{11},
S.~Todo\textsuperscript{12},
M.~Troyer\textsuperscript{13},
M.~Wallerberger \textsuperscript{14}, 
P.~Werner\textsuperscript{15} and
T.~M.~R.~Wolf\textsuperscript{7}
}\end{center}

\begin{center}
(for the ALPS collaboration)
\end{center}

\begin{center}
{\bf 1} Univ Toulouse, CNRS, Laboratoire de Physique Th\'eorique, Toulouse, France
\\
{\bf 2} Department of Physics and Engineering, West Chester University, Pennsylvania, USA
\\
{\bf 3} Department of Physics, Northeastern University, Boston, Massachusetts, USA
\\
{\bf 4} Department of Physics, University of Michigan, Ann Arbor, Michigan, USA
\\
{\bf 5} Department of Physics, University of Warsaw, Warsaw, Poland
\\
{\bf 6} Department of Physics and Physical Oceanography, Memorial University of Newfoundland, St.~John's, Newfoundland, Canada
\\
{\bf 7} Department of Physics, Virginia Tech, Blacksburg, Virginia, USA
\\
{\bf 8} School of Engineering, Mathematics and Physics, University of Kent, Canterbury, UK
\\
{\bf 9} Arnold Sommerfeld Center for Theoretical Physics, Department of Physics, Ludwig-Maximilians-Universit\"at M\"unchen, Munich, Germany
\\
{\bf 10} Department of Physics, Saitama University, Saitama, Japan
\\
{\bf 11} Department of Physics and Astronomy, Middle Tennessee State University, Murfreesboro, Tennessee, USA
\\
{\bf 12} Department of Physics, University of Tokyo, Tokyo, Japan
\\
{\bf 13} Microsoft Quantum, Redmond, Washington, USA
\\
{\bf 14} TU Wien, Institute of Solid State Physics, Austria
\\
{\bf 15} Department of Physics, University of Fribourg, Switzerland\\[\baselineskip]
$\star$ \href{mailto:scarola@vt.edu}{\small scarola@vt.edu}
\end{center}

\section*{\color{scipostdeepblue}{Abstract}}
\textbf{\boldmath{%
We present release 3.0 of the \alps{} (Algorithms and Libraries for
Physics Simulations) project, an open-source software project to develop
libraries and application programs for the simulation of strongly
correlated quantum lattice models such as quantum magnets, lattice
bosons, and strongly correlated fermion systems. As in previous releases,
development is centered on common data formats, on libraries to simplify
and speed up code development, and on full-featured simulation programs
that let non-experts carry out serial or parallel numerical simulations
using the important algorithms for quantum lattice models: classical and
quantum Monte Carlo (QMC) using non-local updates, extended-ensemble
simulations, exact and full diagonalization (ED), the density matrix
renormalization group (DMRG),
and continuous-time QMC solvers for dynamical mean-field theory (DMFT).
Major changes in release 3.0 include distribution of the \pyalps{}
binary through the Python Package Index (\code{pip install pyalps}) and
through Spack for HPC systems; migration of development to GitHub with
continuous integration and automated testing; relicensing of the package
under the permissive MIT license; a completely rebuilt documentation and
tutorial website, including a set of Jupyter-notebook tutorials and
localized content; a broad modernization of the C++
codebase (C++17 compliance, Boost and NumPy~2.0 compatibility, and
warning and dead-code cleanup) together with a major DMRG update and
associated reliability and build-compatibility fixes; the removal of
legacy components (the
VisTrails provenance integration and the TEBD, MPS, and
directed-worm-algorithm application codes); archival of release~3.0.0 with a
Zenodo DOI (\href{https://doi.org/10.5281/zenodo.22775899}{10.5281/zenodo.22775899}); and a
formal governance and sustainability model developed under the US National
Science Foundation (NSF) POSE program.
The software is available from our web server at
\url{https://alps.comp-phys.org}.
}}

\vspace{\baselineskip}

\noindent\textcolor{white!90!black}{%
\fbox{\parbox{0.975\linewidth}{%
\textcolor{white!40!black}{\begin{tabular}{@{}l@{}r@{}}%
  \begin{minipage}{0.6\linewidth}%
    {\small Copyright attribution to authors. \newline
    This work is a submission to SciPost Physics Codebases. \newline
    License information to appear upon publication. \newline
    Publication information to appear upon publication.}
  \end{minipage} & \begin{minipage}{0.4\linewidth}
    {\small Received Date \newline Accepted Date \newline Published Date}%
  \end{minipage}
\end{tabular}}
}}
}


\vspace{10pt}
\noindent\rule{\textwidth}{1pt}
\tableofcontents
\noindent\rule{\textwidth}{1pt}
\vspace{10pt}

\section{Introduction}
\label{sec:intro}

The \alps{} project is dedicated to the simulation of quantum many-body
systems. This paper presents release~3.0 of the \alps{} project and updates the
publications describing the previous releases~\cite{alps10,alps13,alps20}.

Strong interactions between quantum particles, quantum fluctuations,
and competing energy scales give rise to collective phases that are often out
of reach of controlled analytical approximations. The range of experimental
platforms hosting such strongly correlated physics has grown considerably over
the years: condensed-matter systems (frustrated magnets, high-temperature
superconductors, fractional quantum Hall states), atomic and optical systems
(cold atoms in optical lattices), and, increasingly, quantum computers and simulators. Numerical simulations are essential to capture this
physics, and a wide array of computational techniques has been developed to
tame the exponential complexity of the underlying quantum many-body problem.

The increasing reach of these methods comes with a corresponding software
challenge. Modern algorithms combine specialized update schemes, statistical
analysis, sparse linear algebra, parallel execution, and portable data
management. Reimplementing all of these components for each scientific
project duplicates effort and makes validation, reproduction, and long-term
maintenance harder. Shared libraries, stable file formats, tested reference
applications, and worked examples let researchers concentrate on the science
while retaining full access to the implementation and its assumptions.

\alps{} answers this challenge by providing an open-source collection of reusable C++
libraries, Python analysis tools, and application programs for lattice
models. It addresses three characteristic needs of the broad community
working on strongly correlated systems.
(i)~No single computational method can solve all strongly correlated models,
and physical understanding often emerges from combining the results of
several techniques. \alps{} implements a set of state-of-the-art methods that
can be applied to the same model within a common workflow and with common
file formats.
(ii)~Many researchers need reference numerical data, for instance to compare
with experiments, without having to master the underlying algorithms. The
extensively tested \alps{} applications can be used as black boxes with
straightforward input and output.
(iii)~Conversely, researchers developing new or improved numerical methods
want to focus on the novel parts of their algorithms. The \alps{} libraries
and tools provide the underlying infrastructure, making software development
for strongly correlated systems fast and reliable.

Previous versions of \alps{} have been used by hundreds of
researchers in at least 52 countries, have underpinned thousands of
publications, and have been applied in more than 20 research disciplines.  Release~3.0 is both a continuation of the scientific interfaces
documented for release~2.0 and a renewal of the project: the codebase has
been adapted to current toolchains, distribution and testing have been
automated, obsolete components have been retired, and project governance
has been made explicit. The principal release changes are summarized in
Sec.~\ref{sec:new}. Sections~\ref{sec:install}--\ref{sec:applications}
describe installation, common infrastructure, and the application suite;
Secs.~\ref{sec:tutorials}--\ref{sec:project-org} cover documentation,
software quality, reproducibility, governance, licensing, and citation.

\section{The \alps{} project}
\label{sec:project}
The \alps{} project provides an open-source computational framework for the simulation and analysis of quantum and classical lattice models. Rather than centering numerical work on isolated, algorithm-specific implementations, \alps{} organizes commonly needed functionality into shared libraries, data formats, analysis tools, and simulation applications. This approach allows different numerical methods to operate within a common computational environment while reducing duplication in code development and scientific workflows. A central feature of \alps{} is the use of common descriptions and data structures across its libraries and applications. Simulation parameters, lattices, models, and measurements can be specified in forms that are shared across different parts of the software. The lattice and model libraries provide tools for constructing graphs, basis sets, operators, and Hamiltonians, while individual applications implement the numerical algorithms used to study the specified quantum or classical lattice model. Simulation results are stored using portable XML and HDF5 formats and can be processed and analyzed with the Python-based \pyalps{} tools. To support the simulation and analysis of quantum and classical lattice models, the \alps{} project provides:

\begin{itemize} 
\item \textbf{Common data formats.} \alps{} uses standardized, portable data formats to support the exchange, storage, and analysis of simulation data across applications. XML provides human-readable descriptions of parameters, lattices, models, and job structure, while HDF5 provides portable binary storage of simulation results and checkpoints.

\item \textbf{Reusable libraries.} The \alps{} libraries provide common infrastructure for quantum and classical lattice simulations, including parameter handling, XML and HDF5 input/output, symbolic-expression evaluation, serialization and checkpointing, lattice and model construction, statistical analysis of Monte Carlo observables, random-number and numerical utilities, and facilities for serial and parallel execution.

\item \textbf{Python-based evaluation and analysis tools.} \pyalps{} provides the connection between the compiled simulation applications and Python-based analysis workflows. It supports preparation of parameter sets and job files, execution of applications, loading and organizing simulation results, statistical analysis, post-processing, and visualization. 

\item \textbf{Simulation applications.} \alps{} provides ready-to-use implementations of major numerical methods for quantum lattice systems. The release~3.0 application suite includes exact sparse and full diagonalizations, classical Monte Carlo simulations, quantum Monte Carlo methods, density matrix renormalization group calculations, and dynamical mean-field theory with quantum impurity solvers.

\item \textbf{Build, installation, and distribution infrastructure.} \alps{} supports installation through prebuilt \pyalps{} packages distributed through PyPI, source builds using CMake, and Spack packages for HPC environments. The build system supports modern C++ toolchains and both serial and MPI-enabled calculations, providing installation paths appropriate for individual users, developers, and high-performance computing systems. 

\item \textbf{Documentation and tutorials.} \alps{} provides extensive documentation, worked examples, and tutorials designed for students, researchers, and developers. Release 3.0 includes a rebuilt documentation website, structured tutorial tracks, Jupyter-notebook examples, code-development guides, and localized content. 

\item \textbf{Reproducibility and software quality.} Development is organized through GitHub with continuous integration, automated builds, and regression testing. The source code and tutorial material for release~3.0.0 are preserved in a Zenodo archive (\href{https://doi.org/10.5281/zenodo.22775899}{10.5281/zenodo.22775899}), providing a fixed version of the software and examples for future use.

\item \textbf{Open-source licensing and scientific credit.} Beginning with release 3.0, the \alps{} libraries and applications are distributed under the permissive MIT license. The legal terms for software reuse are separated from the project's citation policy, which requests appropriate credit for the original algorithms, their \alps{} implementations, and the shared \alps{} infrastructure.

\item \textbf{Governance and sustainability.} \alps{} maintains a formal governance and sustainability framework developed under the NSF POSE program. Maintainer groups, core maintainers, a governing council, and an advisory board provide mechanisms for code stewardship, scientific validation, project planning, contribution review, and long-term maintenance.

\item \textbf{Community support and outreach.} \alps{} supports a broader user and developer community through its project website \url{https://alps.comp-phys.org}, GitHub repository, mailing lists, Discord server, YouTube resources, workshops, and educational activities. These channels provide user support, disseminate documentation and training materials, and create pathways for students and researchers to participate in the project. 

\end{itemize}

Together, these components provide a common workflow from model definition and numerical simulation to data storage and analysis. For users, the application suite provides implementations of established numerical methods that can be applied through a consistent set of model, data, and analysis tools. For developers, the common libraries provide reusable infrastructure for building and maintaining computational applications without independently implementing functionality such as parameter handling, data storage, scheduling, checkpointing, and statistical analysis.

\alps{} release 3.0 retains this overall framework while modernizing the software and the way it is distributed, maintained, documented, and developed. In addition to updates to the codebase and application suite, the release introduces modern distribution mechanisms, automated testing and continuous integration, updated documentation and tutorials, a permissive MIT license, and a formal governance and sustainability model. The specific technical and organizational changes introduced in release 3.0 are described in Sec.~\ref{sec:new}.

\paragraph{Project history and ALPSCore.}
Development of the original \alps{} codebase slowed down after release~2.0 and
eventually became dormant. During that period, its core libraries were
forked and modernized as the independently maintained ALPSCore
project~\cite{alpscore,ALPSCore2}. Development of the full \alps{} application suite
restarted in 2024 under the \code{ALPSim} GitHub organization, leading to
the 2.3.x maintenance releases and now, with support from the NSF POSE
program, release~3.0. ALPSCore reintegration is planned for release~3.1.

\section{Changes in release 3.0}
\label{sec:new}

Release 3.0 combines modernization of the ALPS software infrastructure with updates to the application suite, documentation, distribution, and project organization. It also removes legacy components that are no longer maintained as part of the current release. Table~\ref{tab:changes} summarizes the major changes and points to where each is described in detail. New features and fixes are discussed in Sec.~\ref{sec:newfeatures},
while removed components and corresponding migration guidance are
described in Sec.~\ref{sec:removed}.

\begin{table}[H]
\centering
\small
\renewcommand{\arraystretch}{1.2}
\begin{tabular}{@{}p{0.72\linewidth}l@{}}
\toprule
\textbf{Change} & \textbf{Details} \\
\midrule
Binary distribution via \code{pip}/PyPI; Spack packages for HPC &
  Sec.~\ref{sec:install} \\
Development on GitHub: CI, automated wheels, regression tests &
  Sec.~\ref{sec:infra} \\
Relicensed under the permissive MIT license, with citation request &
  Secs.~\ref{sec:license}, \ref{sec:citation} \\
Rebuilt website, documentation, and tutorials; localization &
  Sec.~\ref{sec:tutorials} \\
Toolchain modernization: C++17 (selectable to C++23); CMake 3.22+,
  system Boost, and NumPy~2.0 supported & Secs.~\ref{sec:libraries}, \ref{sec:infra} \\
Major DMRG update; QMC and DMFT fixes &
  Sec.~\ref{sec:applications} \\
Removal of legacy components (VisTrails, TEBD, MPS, \code{dwa}) &
  Sec.~\ref{sec:removed} \\
Release source archive with a Zenodo DOI (\href{https://doi.org/10.5281/zenodo.22775899}{v3.0 DOI}) &
  Sec.~\ref{sec:repro} \\
Governance and sustainability model  &
  Sec.~\ref{sec:governance} \\
\bottomrule
\end{tabular}
\caption{Major changes in release 3.0 and where they are described.}
\label{tab:changes}
\end{table}

\subsection{New features and fixes}
\label{sec:newfeatures}

The specific new features and changes in release~3.0 are:
\begin{itemize}
  \item \textbf{Distribution via \code{pip}/PyPI and Spack:} the
        prebuilt \pyalps{} binary installs with
        \code{pip install pyalps} (supported on Python~3.9 and newer); Spack packages
        cover source/HPC installation on Linux and macOS
        (Sec.~\ref{sec:install}).
  \item \textbf{Migration to GitHub} (\url{https://github.com/ALPSim/ALPS})
        with continuous integration, automated wheel builds, and
        regression testing (Sec.~\ref{sec:infra}).
  \item \textbf{MIT relicensing} of libraries and applications
        (Secs.~\ref{sec:license}, \ref{sec:citation}).
  \item \textbf{Rebuilt website, documentation, and tutorials},
        including how to set up lattices and models, Jupyter-notebook
        tutorials, a ``which code to
        choose'' QMC guide, and localized content
        (Sec.~\ref{sec:tutorials}).
  \item \textbf{In-browser documentation assistant} on the project website,
      answering common questions and linking into the documentation.
  \item \textbf{Codebase modernization:} C++17 compliance, fixes for Boost
        deprecation warnings and system-Boost support, NumPy~2.0 compatibility,
        and extensive warning/dead-code cleanup across the libraries
        (Secs.~\ref{sec:libraries}, \ref{sec:infra}).
  \item \textbf{Code updates and fixes:} a major DMRG update with new
        optimizations and methods, plus a crash fix in
        temporary-file creation;
        the directed-loop SSE application corrects the selection of
        the diagonal shift \code{EPSILON};
        five latent bugs fixed in the DMFT QMC driver with regression
        tests; a
        maxent frequency-grid fix and a \code{kron()} stride
        fix in the numeric library. Details are given in
        Sec.~\ref{sec:applications}.
        
  \item \textbf{Governance and sustainability} model established under the
        NSF POSE program (Sec.~\ref{sec:governance}).
\end{itemize}

\subsection{Removed components and migration}
\label{sec:removed}
Release~3.0 removes, rather than deprecates, several components of the
2.x series. Their sources remain in the repository history (see below),
but they are not built, tested, or distributed with release~3.0. For each
removed component the list below states what it was and where existing
workflows go.
\begin{itemize}
  \item \textbf{VisTrails}: the graphical
        workflow and provenance system of release~2.0~\cite{alps20}.
        Simulations are now set up, run, and evaluated with \pyalps{}
        scripts and Jupyter notebooks, and the tutorials ship in that form
        (Sec.~\ref{sec:tutorials}). Existing VisTrails workflows have to be
        re-expressed as \pyalps{} scripts. The release source code and
        included tutorial material are preserved in the software archive
        (Sec.~\ref{sec:repro}).
  \item \textbf{TEBD and MPS}: the two tensor-network
        codes beyond \code{dmrg}, namely the Fortran time-evolving
        block-decimation code for real- and imaginary-time evolution of
        one-dimensional models~\cite{alps20} and the matrix-product-state
        suite (\code{mps\_optim}, \code{mps\_evolve}, \code{mps\_overlap},
        and related tools)~\cite{dolfi2014}, together with their \pyalps{}
        helpers and tutorials. Ground-state calculations of one-dimensional
        models move to the retained \code{dmrg} application
        (Sec.~\ref{sec:applications}). Time evolution has no replacement
        within \alps{}~3.0 but will be replaced in 3.1 with time-dependent \code{dmrg}.
  \item \textbf{Directed worm algorithm (\code{dwa})}: a
        worm-type QMC code for Bose--Hubbard models developed for bosons in
        optical lattices, together with its \pyalps{} helpers for
        band-structure parameters and trap density profiles. It already
        carried a run-time deprecation warning in the 2.3.4 pre-releases.
        Bose--Hubbard models are covered by the retained \code{worm}
        application, including soft-core bosons (tutorial MC-05), and
        hard-core bosons also by \code{dirloop\_sse}; the website's
        code-selection guide compares the two. The optical-lattice and
        trap-specific helpers have no replacement. The other QMC
        applications (\code{looper}, \code{dirloop\_sse}, \code{worm},
        \code{qwl}) remain included.
  \item \textbf{Miscellaneous:} the unmaintained wxPython/VTK lattice-preview
        GUI, four opt-in compatibility headers deprecated
        since 2010, and the two obsolete documentation
        trees \code{doc/} and \code{docs/}, superseded by the
        website.
\end{itemize}

All removed sources remain in the history of the \code{ALPSim/ALPS}
repository. The last 2.x release,
\href{https://github.com/ALPSim/ALPS/tree/b7c08f2e47675e7f9a7733b456da840e4a94e9e4}{\code{v2.3.4-beta.2}}
(February 2026), still contains VisTrails, TEBD, MPS, and \code{dwa}.
Reproducing a calculation that depends on them means building that pinned
revision with period-appropriate dependencies and keeping the original
inputs and analysis scripts. The collaboration does not maintain, test, or
support these components against current toolchains.

\section{Installation and first steps}
\label{sec:install}
\subsection{Installation paths}
\alps{} supports three installation paths on Linux and macOS:
\begin{itemize}
  \item \textbf{Binary (\code{pip}):} \code{pip install pyalps}. Fastest,
        no compilation, no admin privileges. Prebuilt wheels are provided
        for Linux and macOS for CPython 3.9--3.14. The binary build is
        serial; for parallel (MPI) runs use the source or Spack install.
  \item \textbf{Source (CMake):} for developers and users who need custom
        BLAS/LAPACK or MPI parallelism, or who want to modify the code.
        Supported build environments: CMake~3.22 and newer, Boost
        sources~1.76 and newer, BLAS/LAPACK, HDF5, MPI, Python~3.9 and
        newer, and a C++17-capable compiler (currently tested in CI with
        GCC~11--15 and Clang~14--22).
  \item \textbf{Spack:} recommended for HPC operators and cluster users
        whose systems already manage software with Spack. The \code{alps}
        package is part of the Spack builtin repository, so
        \code{spack install alps} needs no additional package repository;
        MPI support is enabled by default. Users without root access can
        also use Spack to install the package in their home directory.
\end{itemize}
On Windows, \alps{} is supported through the Windows Subsystem for Linux
(WSL); a walkthrough is linked from the installation page of the website
(\url{https://alps.comp-phys.org}).

\subsection{A first simulation}
\label{sec:firstrun}
\begin{figure}[t]
  \centering
  \includegraphics[width=0.62\textwidth]{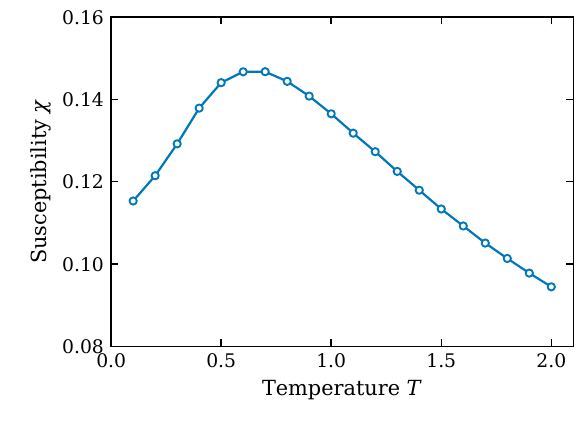}
  \caption{Uniform susceptibility $\chi$ of the $S=1/2$
  antiferromagnetic Heisenberg chain with $L=60$ sites as a function of
  temperature $T$, obtained with the \code{loop} code using the script in
  Sec.~\ref{sec:firstrun}. Statistical errors are smaller than the plot markers.}
  \label{fig:firstrun}
\end{figure}

Using the \alps{} binary install, a complete first calculation for the uniform
susceptibility of a quantum Heisenberg spin chain takes only a few lines of code:
\begin{lstlisting}[language=Python]
import pyalps
import pyalps.plot
parms = [{'LATTICE': "chain lattice", 'MODEL': "spin", 'local_S': 0.5,
          'L': 60, 'J': 1, 'T': t, 'THERMALIZATION': 5000, 'SWEEPS': 50000,
          'ALGORITHM': "loop"}
         for t in [round(0.1 * i, 1) for i in range(1, 21)]]
input_file = pyalps.writeInputFiles('parm', parms)
pyalps.runApplication('loop', input_file)
data = pyalps.loadMeasurements(
           pyalps.getResultFiles(prefix='parm'), 'Susceptibility')
chi = pyalps.collectXY(data, x='T', y='Susceptibility')
chi[0].props['ylabel'] = r'Susceptibility $\chi$'
chi[0].props['xlabel'] = r'Temperature $T$'
pyalps.plot.plot(chi)
\end{lstlisting}
The script writes the \alps{} input files, runs one task per temperature
with the \code{loop} continuous-time quantum Monte Carlo code, loads the
measurements, and plots the susceptibility. The twenty tasks complete in
seconds on a laptop; the result is shown in
Fig.~\ref{fig:firstrun}.
New users are pointed to the ``Which code to choose for your
calculation'' guide, the
\code{crash\_\allowbreak course\_\allowbreak pyalps} notebook, and the
transverse-field Ising and spin-gap notebooks
(Sec.~\ref{sec:tutorials}) as first steps.
\section{Inside \alps{}: data formats, libraries, and Python tools}
\label{sec:inside}

\subsection{Data formats}
\label{sec:formats}
Common data formats allow the same model and result files to move between
applications and analysis tools. \alps{} continues to use XML for
human-readable descriptions of parameters, lattices, models, and job
structure, and HDF5 for portable binary storage of simulation results and
checkpoints. 
\pyalps{}
can load the resulting measurements, spectra, eigenstate data, and simulation
properties without exposing the storage details to routine analysis code.

The XML schemas and the layered HDF5 interface were introduced in earlier
\alps{} releases and are described in detail in
Ref.~\cite{alps20}. Release~3.0 continues to use the established XML and HDF5 data
interfaces. Current format
documentation and examples are maintained on the project website. The C++
HDF5 library remains usable independently of the application programs.

\subsection{Libraries}
\label{sec:libraries}
The \alps{} libraries provide the common infrastructure used by the
applications and by external codes built on the project:
\begin{itemize}
  \item XML parsing and output, symbolic-expression evaluation, and
        parameter handling;
  \item HDF5 serialization and the \code{osiris} checkpoint-serialization
        layer;
  \item scheduler and \code{parapack} facilities~\cite{troyer1998scheduler}
        for serial and MPI execution, load distribution, and checkpointing;
  \item lattice and model libraries for constructing graphs, basis sets,
        operators, and Hamiltonians from common descriptions;
  \item \code{alea} tools for recording and statistically evaluating
        Monte Carlo observables; and
  \item random-number, numerical, and Boost Graph Library extension
        utilities used across the simulation codes.
\end{itemize}
These interfaces preserve the separation emphasized in release~2.0:
scientific applications supply the algorithm, while common concerns such
as input, scheduling, measurements, and persistence are handled once.
Release~3.0 modernizes the implementation and build environment around
those interfaces. CMake~3.22 and newer are supported, C++17 is the default
language mode with C++20 and C++23 exercised in CI, and the supported
matrix covers Boost~1.76 through 1.91 together with the HDF5 libraries
provided on the tested Linux and macOS platforms. The Boost numeric
bindings were also moved under \code{bindings/} and treated as system
headers, isolating third-party compatibility warnings from project
sources.

\subsection{Evaluation tools in Python (\pyalps{})}
\label{sec:pyalps}
\pyalps{} connects the compiled applications to an interactive analysis
workflow. It writes parameter sets and job files, launches applications
locally or through MPI, locates their outputs, and loads measurements from
the common XML/HDF5 formats. Results are represented as data sets carrying
both numerical arrays and descriptive properties, so parameter sweeps can
be collected, filtered, and plotted with metadata intact. The package also
exposes statistical-analysis facilities and plotting through the standard
Python scientific stack.

The package is organized into modules with distinct roles,
summarized in Table~\ref{tab:pyalps}.

\begin{table}[H]
\centering
\small
\renewcommand{\arraystretch}{1.2}
\begin{tabular}{@{}p{0.30\linewidth}p{0.62\linewidth}@{}}
\toprule
\textbf{Module} & \textbf{Role} \\
\midrule
\code{pyalps.tools} &
  Prepares parameter sets and job files, launches applications, and
  collects results into data sets \\
\code{pyalps.load} &
  Reads measurements, spectra, eigenstate data, iteration histories, and
  simulation properties from the HDF5 output; XML file names are accepted
  and mapped to their HDF5 counterparts \\
\code{pyalps.dataset} &
  Provides the \code{DataSet} and \code{ResultFile} containers that carry
  numerical arrays together with the parameters that produced them \\
\code{pyalps.hlist} &
  Hierarchical-list utilities for the nested structures that the loaders
  return, including \code{flatten}, \code{deep\_flatten}, \code{hmap},
  and \code{depth} \\ 
\code{pyalps.alea}, \code{pyalps.math} &
  Monte Carlo statistics, autocorrelation analysis, and error propagation
  on measured quantities \\
\code{pyalps.plot}, \code{pyalps.plot\_core} &
  Render results through Matplotlib and export them to text, Grace, and
  gnuplot \\
\code{pyalps.hdf5} &
  Exposes the archive interface directly \\
\code{pyalps.fit\_wrapper} &
  Least-squares fitting \\
\code{pyalps.lattice} &
  Visualizes lattice structures \\
\code{pyalps.apptest} &
  Drives the application regression tests \\
\code{pyalps.pytools}, \code{pyalps.cxx} &
  Compiled bindings to the \alps{} C++ libraries \\
\bottomrule
\end{tabular}
\caption{\pyalps{} modules and their roles in release~3.0. Further modules (on impurity solvers and analytic-continuation) are described in the text.}
\label{tab:pyalps}
\end{table}

Beyond driving the standalone executables, \pyalps{} also ships compiled
extension modules built from the application sources themselves. These are
installed when LAPACK is present and the application suite is built, which
is the configuration used for the published wheels. Two of them wrap the
continuous-time impurity solvers: \code{pyalps.cthyb} for the
hybridization expansion and \code{pyalps.ctint} for the interaction
expansion, each exposing a \code{solve()} entry point. A third,
\code{pyalps.maxent}, wraps the maxent analytic-continuation tool and
exposes \code{AnalyticContinuation}.

A DMFT self-consistency loop can therefore be written entirely in Python,
with the impurity solver called in-process rather than as a subprocess;
the \code{hybridization-01} to \code{hybridization-04} tutorials are built
this way. The scheduler-driven route through \code{runDMFT} and the
\code{dmft} executable remains available and is what the DMFT notebook
tutorials use.

Python is now the sole supported evaluation path in \alps{}. The
VisTrails integration described in Ref.~\cite{alps20} has been removed;
reproducible examples are supplied as Jupyter notebooks instead
(Sec.~\ref{sec:repro}). Application-specific \pyalps{} support for the
removed TEBD and MPS codes was deleted with those applications, as was the helper module for the directed worm
application when that code was removed. The general data,
execution, evaluation, and plotting interfaces remain available for the
applications shipped in release~3.0.

\section{Applications}
\label{sec:applications}
Table~\ref{tab:apps} gives an overview of the application codes shipped
with release~3.0; the subsections below describe the status of and
changes to each. Section~\ref{sec:citation} discusses how to cite these codes using the citation matrix
available on the \alps{} website. 

\begin{table}[H]
\centering
\footnotesize
\setlength{\tabcolsep}{3.5pt}
\begin{tabular}{@{}l>{\raggedright\arraybackslash}p{0.24\linewidth}>{\raggedright\arraybackslash}p{0.10\linewidth}>{\raggedright\arraybackslash}p{0.12\linewidth}>{\raggedright\arraybackslash}p{0.16\linewidth}@{}}
\toprule
\textbf{Code} & \textbf{Method and typical use} &
\textbf{Release 3.0 note} & \textbf{Algorithm Citation} & \textbf{Implementation Citation} \\
\midrule
\code{scheduler} & parallel Monte Carlo scheduler and job management &
  retained & N/A & \cite{troyer1998scheduler} \\
\code{sparsediag} & sparse ED (Lanczos); ground and low-lying states &
  retained & \cite{lanczos1950} & This Work \\
\code{fulldiag} & full ED; spectra and thermodynamics &
  retained & N/A & This Work \\
\code{spinmc} & classical MC with local and cluster updates &
  retained & \cite{swendsenwang1987,wolff1989} & This Work \\
\code{looper} & loop-cluster QMC for unfrustrated quantum magnets &
  citation notice & \cite{evertz1993,beard1996,evertz2003} & \cite{todo2001} \\
\code{dirloop\_sse} & directed-loop SSE for quantum magnets and bosons &
  ergodicity fix & \cite{sandvik1999,syljuasen2002} & \cite{alet2005directed} \\
\code{worm} & worm-algorithm QMC for lattice bosons &
  retained & \cite{prokofev1998,Prokofev1998JETP} & This Work \\
\code{qwl} & quantum Wang--Landau sampling over a broad temperature range &
  citation notice & \cite{wang2001prl,wang2001pre,troyer2003qwl} & This Work \\
\code{dmrg} & DMRG for one- and quasi-one-dimensional ground states &
  update & \cite{white1992,white1993,schollwock2005,hallberg2006} & \cite{feiguin2013} \\
\code{dmft} & CT-INT, CT-HYB, Hirsch--Fye, and DMFT self-consistency &
  fixes and citation notices & \cite{rubtsov2005,werner2006,gull2011} &
  \cite{gull2011cpc} \\
\bottomrule
\end{tabular}
\caption{Application codes in the release build and their principal
release-3.0 changes. Relevant references for each algorithm and their implementation are listed here and are maintained on the project website.  Sec.~\ref{sec:citation} discusses citation instructions.}
\label{tab:apps}
\end{table}

\subsection{Exact and full diagonalization (\code{sparsediag}, \code{fulldiag})%
}
The diagonalization applications favor a common model interface and
flexible measurements over method-specific optimization for the largest
possible Hilbert spaces. \code{sparsediag} uses the Lanczos
method~\cite{lanczos1950} to calculate a requested number of low-lying
eigenstates, making it suitable for ground-state properties and low-energy
spectra. \code{fulldiag} computes the complete spectrum with dense
linear-algebra routines and can therefore evaluate thermodynamic
quantities as functions of temperature. Both applications use conserved
quantum numbers and, where available, translation symmetry to decompose
the Hilbert space into sectors. Users can request averaged or local
site- and bond-operator measurements and correlation functions through
the shared model and measurement descriptions. Release~3.0 retains both
applications.

\subsection{Classical Monte Carlo (\code{spinmc})}
\code{spinmc} provides local and cluster Monte Carlo updates for classical
Ising, XY, Heisenberg, O(4), and Potts models. Single-cluster updates of
Wolff type~\cite{wolff1989}, building on the cluster formulation of
Swendsen and Wang~\cite{swendsenwang1987}, reduce critical slowing down for
unfrustrated models with isotropic couplings and no external field; local
updates cover frustrated systems, anisotropic couplings, and models in a
field, for which cluster updates are rejected. When the update type is left
unspecified, the application selects cluster updates automatically for
unfrustrated lattices. It records energy, magnetization, staggered
magnetization, susceptibility, and model-dependent observables through the
common measurement infrastructure, from which \code{spinmc\_evaluate}
derives the specific heat, connected susceptibility, and Binder cumulants.
Release~3.0 retains the application.

\subsection{Quantum Monte Carlo}
\subsubsection{Loop algorithm (\code{loop})}
\code{loop} implements loop-cluster updates for quantum spin models with arbitrary spin quantum number $S$~\cite{todo2001,evertz1993,beard1996,evertz2003}.
It supports both continuous-time path-integral and stochastic-series-expansion representations, using non-local cluster updates in space and imaginary time.
Supported interactions include $XXZ$ exchange, single-ion anisotropy, and longitudinal and transverse magnetic fields.
The algorithm is best suited for unfrustrated, sign-problem-free quantum-spin models.
Measurements include energy, magnetization, susceptibility, spin correlations, and spin stiffness, with cluster-based improved estimators used where applicable.
Release~3.0 retains the implementation and adds a startup citation notice.

\subsubsection{Directed loop (\code{dirloop\_sse})}
\code{dirloop\_sse} implements operator-loop and directed-loop updates in the
stochastic-series-expansion representation
\cite{sandvik1999,syljuasen2002,alet2005directed}. It supports quantum-spin
models and soft-core bosonic models on arbitrary supported lattices.
Release~3.0 corrects the selection of the diagonal shift \code{EPSILON}.
Note that this shift does not eliminate sign problems arising from off-diagonal
matrix elements in frustrated systems.
\subsubsection{Worm algorithm (\code{worm})}
\label{sec:worm}
\code{worm} implements continuous-time path-integral QMC using worm
updates~\cite{prokofev1998,Prokofev1998JETP}. It supports
sign-problem-free soft-core bosonic models and unfrustrated quantum-spin
models on arbitrary supported lattices. Measurements include
energy and, where applicable, density, stiffness, and Green functions.
Release~3.0 contains no changes compared to the previous version.

\subsubsection{Quantum Wang--Landau (\code{qwl})}

\code{qwl} applies quantum Wang--Landau sampling to the stochastic-series-expansion representation \cite{wang2001prl,wang2001pre,troyer2003qwl,wessel2007qwl}. By performing a random walk in the space of series expansion coefficients, it achieves a flat histogram in their orders. One simulation can yield energy, free energy, entropy, specific heat, and magnetic observables across a range of temperatures. The implementation targets isotropic (anti-)ferromagnetic spin-$1/2$ Heisenberg models on non-frustrated lattices at zero field.
Release~3.0 adds a startup notice identifying the recommended framework
and method citations. 

\subsection{Density matrix renormalization group (\code{dmrg})%
}
The density matrix renormalization group (DMRG) method provides a variational route to
ground states of one-dimensional and quasi-one-dimensional quantum lattice
models~\cite{feiguin2013,white1992,white1993,schollwock2005,hallberg2006}.
The method diagonalizes the Hamiltonian in a reduced basis that has been
systematically optimized to minimize loss of information about the target
state (usually the ground state of the system).
This is done by applying a change of basis
called the density-matrix projection, equivalent to finding a Schmidt basis decomposition for the state using a singular value decomposition---the same matrix decomposition underlying principal component analysis. The algorithm iteratively updates the renormalized block bases during sweeps through the system until convergence is achieved. As a result, one obtains a faithful representation of the ground state of the Hamiltonian in a reduced Hilbert space. The dimension is controlled by the number of retained states, and the discarded weight provides an estimate of the truncation error. In modern tensor-network language, the method produces an approximation to the ground state in the form of a matrix product state (MPS).  The application does not expose an MPS interface to the user in release~3.0, but this feature is planned for future releases.  The application supports ground-state and low-lying excited-state calculations and measures observables through the common ALPS measurement infrastructure.

The DMRG code (\code{dmrg}) received a major update that optimizes
linear-algebra operations and introduces useful functions intended to
facilitate further DMRG-derived applications, together with
a fix for a crash in the temporary-file creation utility (\code{mkstemp}).
The algorithm for building the Hamiltonian has been optimized, and the
calculation of two-point correlation functions has been greatly improved
by evaluating expectation values when operators are on separate blocks,
also significantly reducing memory and storage requirements.
The DMRG
tutorials were renumbered and synchronized with the rebuilt website.

\subsection{Dynamical mean-field framework}

The DMFT code consists of a self-consistency framework for single-site
dynamical mean-field theory~\cite{metzner1989,georges1992,georges1996} as well as
interaction-expansion (CT-INT)~\cite{rubtsov2005} and
hybridization-expansion (CT-HYB)~\cite{werner2006} quantum impurity
solvers; implementation and method overviews are given in
Refs.~\cite{gull2011cpc,gull2011}. A legacy impurity solver that implements
the Hirsch--Fye algorithm~\cite{hirschfye1986} is also provided as a
reference. The DMFT framework is accompanied by tutorials on the
metal-to-insulator transition.
Release 3.0 fixes five latent bugs in the DMFT QMC driver, now
covered by regression tests; the maxent analytic-continuation tool received a
frequency-grid fix.

\section{Tutorials, documentation, and community}
\label{sec:tutorials}
The project website has been rebuilt and now organizes an extensive,
pedagogically structured tutorial set, summarized in
Table~\ref{tab:tutorials}.

\begin{table}[H]
\centering
\small
\renewcommand{\arraystretch}{1.2}
\begin{tabular}{@{}lp{0.18\linewidth}p{0.48\linewidth}@{}}
\toprule
\textbf{Track} & \textbf{Tutorials} & \textbf{Coverage} \\
\midrule
Monte Carlo & guide + MC-01--MC-09 & ``which code to choose'';
  autocorrelations and equilibration; susceptibilities; magnetization;
  measurements; bosons; quantum Wang--Landau; classical and quantum
  phase transitions; quantum Monte Carlo \\
Exact diagonalization & ED-01--ED-06 & sparse diagonalization, gaps,
  spectra, criticality, phase transition, full diagonalization \\
DMRG & DMRG-01--DMRG-06 & introduction, Heisenberg spin chains,
  ground-state energies, gaps, local observables, correlations \\
 & DMRG-07--DMRG-11 &
  spinless fermions: introduction/Jordan--Wigner
  transformation, ground-state energies, particle-number sectors, model
  types in 1D, open vs.\ closed boundary conditions \\
DMFT & DMFT-01--DMFT-09 & introduction, hybridization, interaction,
  Mott and orbital-selective Mott transitions, paramagnetic and antiferromagnetic solutions,
  Hirsch--Fye, lattices, N\'eel transition \\
Lattices and models & LM-01, LM-02 & honeycomb antiferromagnet; custom
  model Hamiltonians \\
Jupyter notebooks & 10 notebooks & transverse-field quantum Ising model;
  spin gap and spectra of one-dimensional quantum systems; ground-state
  energies and gap extrapolation for spin-1/2 and spin-1 chains; DMFT of
  the Hubbard model on a Bethe lattice; energy spectra of qubits \\
\bottomrule
\end{tabular}
\caption{Tutorial tracks on the rebuilt project website.}
\label{tab:tutorials}
\end{table}
The documentation additionally covers models (Ising, transverse-field
Ising, Heisenberg, spinless fermion, Hubbard, $t$--$J$, hard-core boson,
Bose--Hubbard, Kondo lattice), methods, an API reference, the ALPS
libraries, code-development guides (Code-00--Code-04), and ``ALPSize your
codes'' guides.
The tutorials are continually being expanded and updated to broaden their reach and improve their clarity.
In particular, the website has been restructured with these goals in mind, and its English content has been translated into Chinese and Japanese to reach a wider user base. The website also carries a built-in documentation assistant: a search widget that answers common questions about installation, models, applications, and tutorials, and links directly to the relevant page. It runs entirely in the reader's browser against a curated question bank and an index generated from the documentation at build time, so it requires no backend service or API key and sends no query data anywhere.
Community support is available through the GitHub repository (\url{https://github.com/ALPSim/ALPS}), a dedicated Discord server (\url{https://discord.gg/JRNWnnva9g}), and a YouTube channel with demonstration and help videos (\url{https://youtube.com/@ALPS-Collaboration}).
\begin{figure}[t]
  \centering
  \includegraphics[width=\linewidth]{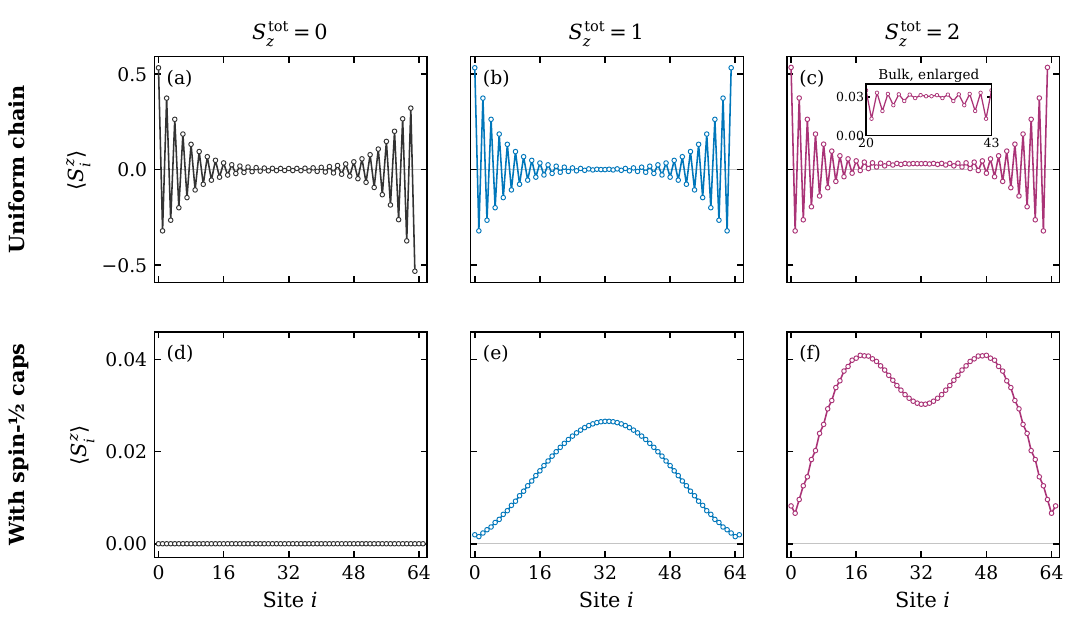}
  \caption{Local magnetization $\langle S_i^z \rangle$ in the spin-1
    Heisenberg chain with open boundaries, resolved by total magnetization.
    (a--c) $64$ spin-1 sites. In sector~0, DMRG selects an
    opposite-edge-polarized combination of the nearly degenerate singlet
    and $S_z^{\mathrm{tot}}=0$ triplet; the exact finite-chain singlet
    would instead have zero local magnetization. Sector~1 carries the
    edge excitation, while sector~2 adds a bulk magnon (inset: enlarged
    central region).
    (d--f) The same $64$ spin-1 sites with two additional
    spin-$\tfrac{1}{2}$ end caps ($66$ sites total). Screening of the
    edge modes yields a singlet in sector~0 and bulk-magnon profiles in
    sectors~1 and~2. The two rows use different vertical scales.
    Lines connect the measured site values.}
  \label{fig:dmrg05}
\end{figure}

\subsection{Tutorial: DMRG} 
This sample tutorial demonstrates the block diagonalization of the spin-1 (Haldane) chain Hamiltonian using symmetries associated with conserved quantum numbers. It covers how to use the \code{dmrg} application and analyze the system using local observables. 
Because the total magnetization
$S_z^{\mathrm{tot}}$ is conserved, the ground-state search is performed
independently within each magnetization sector, and site-resolved
\emph{local observables} then give a real-space picture of the low-lying
excitations (Fig.~\ref{fig:dmrg05}). The local magnetization profile is
recorded with \code{MEASURE\_LOCAL[Local magnetization] = Sz} on a
$64$-site chain.

Reading the profile $\langle S_i^z \rangle$ sector by sector distinguishes
a boundary excitation from a bulk one:
\begin{itemize}
  \item The $S_z^{\mathrm{tot}} = 1$ state places all of its weight in
        exponentially localized modes at the two chain ends and costs
        essentially no energy, so it is nearly degenerate with the
        $S_z^{\mathrm{tot}} = 0$ sector. This near-degeneracy reflects the
        emergent free spin-$\tfrac{1}{2}$ at each open end.
  \item The $S_z^{\mathrm{tot}} = 2$ state is the first sector that cannot
        be built by reorienting the edge spins alone, because the two
        emergent spin-$\tfrac{1}{2}$ edge modes can contribute at most
        $S_z^{\mathrm{tot}} = 1$. It must therefore carry a genuine bulk
        magnon, visible as additional weight spread through the interior of
        the chain.
\end{itemize}
The tutorial also showcases the utility of open boundary conditions in \alps{} by attaching spin-$\tfrac{1}{2}$ caps to the chain ends (Fig.~\ref{fig:dmrg05}, bottom).
The two extra half-spins Kondo-screen the
emergent edge modes into a unique singlet: the local magnetization at the boundaries drops by more than an order of magnitude, so the ends become the quietest rather than the loudest part of the chain, and sectors~1 and~2 display clean bulk-magnon standing-wave profiles. 
As a sector-consistency check, $\sum_i \langle S_i^z \rangle$
agrees with $S_z^{\mathrm{tot}}$ to within $6\times10^{-7}$ for the
profiles shown. Figure~\ref{fig:dmrg05} uses up to $200$ retained states
and $10$ sweeps; comparison with $100$ states and $6$ sweeps changes
local magnetizations by less than $5\times10^{-6}$ and energies by
less than $2\times10^{-6}J$. These checks establish the stability of the
plotted profiles, without resolving the near-degenerate singlet--triplet
mixing in the uncapped sector~0.

\subsection{Tutorial: DMFT}
This sample tutorial introduces the \code{dmft} self-consistency loop with the
hybridization-expansion continuous-time quantum Monte Carlo impurity solver
(CT-HYB), selected with \code{SOLVER = "hybridization"}. As a benchmark it
reproduces Fig.~11 of the DMFT review~\cite{georges1996}: the half-filled
Hubbard model on the Bethe lattice with $U=3$ and hopping
$t=1/\sqrt{2}$ (bandwidth $W=4t$), which orders antiferromagnetically upon
cooling. A single \pyalps{} script writes the parameter files for inverse
temperatures $\beta=6$ and $12$, runs the
self-consistency, and plots the results. The same calculation is repeated
with the interaction-expansion and Hirsch--Fye solvers in DMFT-03 and
DMFT-07, so the three solvers can be compared directly.

\begin{figure}[t]
 \centering
 \includegraphics[width=\linewidth]{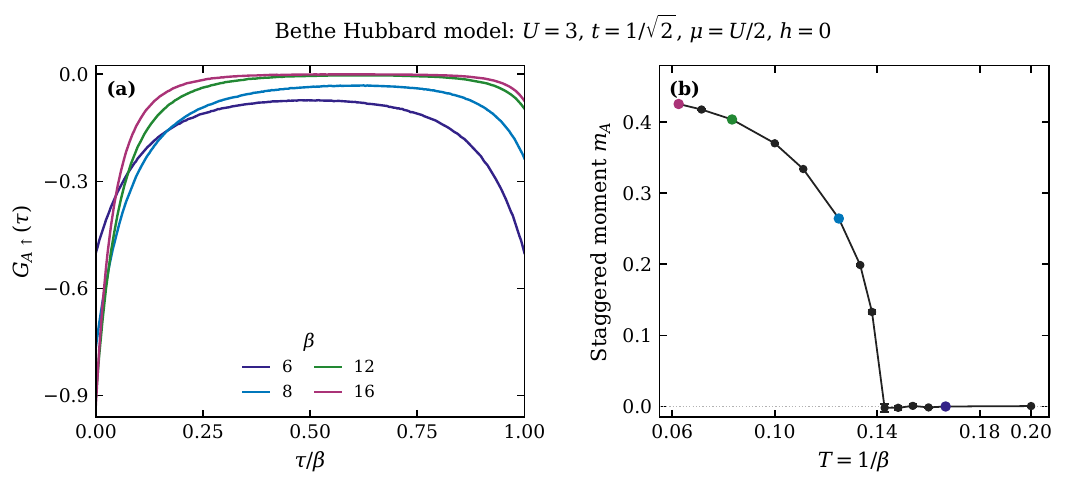}
 \caption{Antiferromagnetic DMFT for the half-filled Hubbard model on the
Bethe lattice, with $U=3$, $t=1/\sqrt{2}$ (half-bandwidth $D=2t$), and zero
physical field, solved with the ALPS CT-HYB impurity solver.
(a) Imaginary-time Green's function for one spin on sublattice $A$ at
$\beta=6,8,12,16$. The opposite spin follows from
$G_{A\downarrow}(\tau)=G_{A\uparrow}(\beta-\tau)$; this exact half-filling
symmetry is also used to average the two measured estimators.
(b) Staggered moment $m_A=(n_{A\downarrow}-n_{A\uparrow})/2$ versus
temperature $T=1/\beta$, sampled more densely near the onset of order.
Colored markers identify the temperatures shown in (a).
All moments use a common sublattice orientation.
The signed moment is used to avoid an absolute-value bias near zero.
Bands and error bars denote one standard error, estimated from sixteen
independently refined DMFT branches and checked against native binned
Monte Carlo errors. Near the transition, longer runs from weakly and strongly
ordered starting baths test independence from initialization.
Lines connect sampled values; no smoothing or transition-temperature fit is
applied.}
 \label{fig:dmft02}
\end{figure}

The tutorial highlights several practical aspects of a DMFT calculation:
\begin{itemize}
  \item \textbf{Symmetry breaking.} With \code{ANTIFERROMAGNET = 1} and
        \code{SYMMETRIZATION = 0}, the self-consistency permits
        antiferromagnetic order and allows the two spin flavors to differ.
        A small initial field \code{H\_INIT} seeds a spin-asymmetric Weiss
        field, helping the iteration reach the ordered solution below the
        transition. The physical field is set to zero.
  \item \textbf{Observables.} The N\'eel transition is visible directly in
        the imaginary-time Green's function $G_f(\tau)$: the spin-up and
        spin-down curves split at low temperature, and the occupation of
        each flavor follows from $n_f = -G_f(\tau=\beta^-)$.
  \item \textbf{Convergence.} The function \code{pyalps.loadDMFTIterations}
        loads iteration-resolved Green's functions. With
        \code{OMEGA\_LOOP = 1}, convergence is assessed from the largest
        change in the interacting Matsubara Green's function across all
        sampled frequencies and spin flavors,
        \begin{equation*}
          \max_{n,f}\left|G_f^{(k+1)}(i\omega_n)-G_f^{(k)}(i\omega_n)\right|
          \leq \code{CONVERGED}.
        \end{equation*}
        The loop also stops after \code{MAX\_IT} iterations, so termination
        alone does not establish convergence. The self-energy, obtained
        from the Dyson equation
        \begin{equation*}
          \Sigma_f(i\omega_n)=\mathcal{G}_{0,f}(i\omega_n)^{-1}
          -G_f(i\omega_n)^{-1},
        \end{equation*}
        provides an additional sensitive diagnostic. Its iteration
        dependence should be assessed together with Monte Carlo noise.
\end{itemize}
Figure~\ref{fig:dmft02} shows higher-statistics CT-HYB calculations for
this model at fourteen temperatures, with denser sampling near the onset of
antiferromagnetic order. Independent DMFT branches estimate uncertainty from
both Monte Carlo sampling and fluctuations of the self-consistent bath.
Longer runs from weakly and strongly ordered starting baths check convergence
near the transition.

\section{Software engineering, quality assurance, and reproducibility}
\label{sec:infra}

\subsection{Continuous integration and packaging}
Development has moved to GitHub (\url{https://github.com/ALPSim/ALPS}).
Continuous integration via GitHub Actions builds and tests the code and produces binary Python wheels on each change (\code{build.yml}, \code{build\_wheels.yml}), and an automated issue-triage workflow and structured issue templates (bug report, feature request, simulation help, website help) support the community. 
Continuous/regression testing is driven through CTest. The build system is CMake (versions 3.22 and newer are supported, with the C++ standard defaulting to C++17 and selectable up to C++23); prebuilt Python wheels have been published to PyPI since October 2024, and a Spack recipe (in the upstream Spack package repository) provides source/HPC installation.

\subsection{Codebase modernization}
A substantial portion of the work since release~2.0 has been a
modernization and portability effort so that \alps{} builds cleanly with
current toolchains: removal of language constructs dropped in C++17
(e.g.\ \code{std::unary\_function}/\allowbreak\code{binary\_function},
\code{bind1st}/\code{bind2nd}, \code{ptr\_fun}), replacement of
interfaces deprecated by current platform SDKs (\code{sprintf} in
favour of the bounds-checked \code{snprintf}), fixes for
Boost deprecation warnings and support for a system-provided Boost,
compatibility with NumPy~2.0, and broad warning and dead-code cleanup
across the \code{alea}, \code{ietl}, \code{scheduler}, \code{parapack},
\code{numeric}, and HDF5 layers. The continuous-integration matrix was
redesigned in 2026 to separate the dimensions that the project supports
and tests. Linux jobs cover Ubuntu~22.04 and 24.04, GCC~11--15,
Clang~14--22, Python~3.9--3.14, and supported Boost releases from
1.76 through 1.91. The default C++17 build is complemented by explicit
C++20 and C++23 jobs. The macOS matrix covers macOS~14, 15, and 26,
including Apple Clang, the macOS~15 Intel runner, and GCC~13/14 builds.
This matrix turns portability claims into continuously checked release
criteria rather than one-time build reports.

\subsection{Software archive and reproducibility}
\label{sec:repro}
The source code for \alps{} release~3.0.0 is archived on Zenodo at
\href{https://doi.org/10.5281/zenodo.22775899}{10.5281/zenodo.22775899}.
The archive includes the tutorial inputs, Python scripts, and Jupyter
notebooks distributed with that release. It preserves a fixed version of
the software and examples that readers can obtain and cite.  Subsequent releases are archived
under the same Zenodo concept record,
\href{https://doi.org/10.5281/zenodo.22775898}{doi:10.5281/zenodo.22775898},
which always resolves to the most recent version.

To make an individual calculation reproducible, users should record the
\alps{} version and software environment and retain the input parameters,
simulation output, and analysis scripts. Together, these specify which
software was used, how the calculation was run, and how the reported
results were obtained.

\section{Governance, license, and citation}
\label{sec:project-org}

\subsection{Governance and sustainability}
\label{sec:governance}

After many years of support from ETH Zurich and the Simons Foundation, \alps{}
is now supported by the US National Science Foundation POSE program. The
project has moved to a hierarchical shared-governance model comprising
application maintainers, core maintainers, a governing council, and an
external advisory board. Maintainer groups set the scope of individual
codes and nominate core maintainers; core maintainers respond to community
issues and certify proposed changes through compilation, testing, and
scientific validation. The governing council appoints maintainers, sets
project roadmaps and dependency policy, makes deprecation decisions, and
leads the publication of release papers. The advisory board provides
independent guidance on technical direction and community development.

Contributions are proposed and reviewed through GitHub under a consensus
process. The documented policy accepts a proposal when the relevant
maintainers agree or when no objection is raised during a review
period; contested decisions may be appealed to the governing council.
New application and library contributors coordinate onboarding with the
council and make a continuing maintenance commitment. A four-level student
onboarding path progresses from installation feedback, through tutorial
development and code maintenance, to contributions of research methods.
Community participation is supported through workshops, GitHub, and Discord. The
governance, contribution, maintainer, and onboarding documents are
published at \url{https://alps.comp-phys.org}; the software archive
and reproducibility guidance are described in Sec.~\ref{sec:repro}.

\subsection{License}
\label{sec:license}
Beginning with release~3.0, the \alps{} libraries and applications are
distributed under the permissive MIT license, replacing the previous
\alps{} library and application licenses. As discussed in Sec.~\ref{sec:citation}, users are still requested to
cite this paper and the relevant algorithm/application papers; the
repository-level guidance is listed in \code{CITATION.md}, and applications
with dedicated notices print relevant recommendations to standard output.

\subsection{Citation policy}
\label{sec:citation}

The MIT license establishes the legal terms for reuse; the citation policy
is a separate request for scientific credit. Citations for work built on
\alps{} fall into three tiers:
\begin{enumerate}
  \item \textbf{The original algorithm.} Many algorithms implemented in
        \alps{} were developed independently of the library. Publications
        using these algorithms should cite the original research that
        introduced them.
  \item \textbf{The implementation.} Individual \alps{} applications were
        contributed by research teams and are often accompanied by
        implementation papers, which should be cited when the
        corresponding application is used.
  \item \textbf{The \alps{} library itself.} This citation credits the
        shared community infrastructure: the base libraries on which the
        applications are built, ongoing code maintenance, and user
        support. Users are asked to cite the most recent release paper.
\end{enumerate}

Table~\ref{tab:apps} lists the relevant algorithm and implementation references as of this release.  The DMRG code illustrates all three citation tiers: the algorithm is credited
by citing the original density-matrix renormalization-group papers~\cite{white1992,white1993,schollwock2005,hallberg2006};
the \code{dmrg} application is credited by citing its implementation
paper~\cite{feiguin2013}; and the framework is credited by citing the
present paper. The GitHub repository contains a master citation file that is reproduced on the project website at
\url{https://alps.comp-phys.org}.  The source code, data files, and website all encourage proper referencing for scientific credit. 

Release~3.0 also moves citation guidance closer to the calculation by
adding or updating startup notices in the DMFT framework and solvers,
\code{qwl}, \code{looper}, and \code{dmrg}. Each notice recommends the framework paper together with the
method-specific references for that application. This mechanism makes the
request visible at the point of use while leaving the software's MIT terms
unchanged.

\section{Conclusions and outlook}
\label{sec:future}
\label{sec:conclusions}

Release~3.0 re-establishes \alps{} as a maintained, citable platform for
strongly correlated lattice-model simulations. It preserves the common
formats, reusable libraries, and reference applications on which existing
workflows depend, while adapting their implementation to current
compilers, dependencies, Python environments, and distribution channels.
Automated testing, release archival, an MIT license, explicit citation
guidance, and shared governance turn that technical renewal into a basis
for continued community maintenance.

Near-term work will concentrate on keeping tutorials and application
interfaces synchronized with releases, extending regression coverage, and
making it easier for contributors to build and test ALPS with supported compilers and dependencies.  Future releases will be archived on Zenodo under the concept DOI \href{https://doi.org/10.5281/zenodo.22775898}{10.5281/zenodo.22775898}.  ALPSCore reintegration is planned for release 3.1.  Additional DMRG features will include the export of MPS states and time-dependent DMRG.  Proposals for upgrades and new contributions will be tested,
discussed, and considered through the governance process described in
Sec.~\ref{sec:governance}, together with other roadmap proposals.
\section*{Acknowledgements}

We thank the following people for useful discussions and contributions to
previous versions of ALPS:
D.~Abrahams,
A.~F.~Albuquerque,
A.~E.~Antipov,
B.~Bauer,
G.~Carcassi,
L.~D.~Carr,
X.~Chen,
P.~Corboz,
P.~Dayal,
Q.~Dong,
H.~G.~Evertz,
J.~Freire,
S.~Fuchs,
A.~Gaenko,
L.~Gamper,
A.~Grzesik,
J.~Gukelberger,
S.~G\"urtler,
A.~Hehn,
A.~Honecker,
R.~Igarashi,
S.~V.~Isakov,
M.~K\"onz,
D.~Koop,
M.~K\"orner,
A.~Kozhevnikov,
A.~L\"auchli,
R.~Levy,
P.~N.~Ma,
S.~R.~Manmana,
P.~Mates,
M.~Matsumoto,
H.~Matsuo,
I.~P.~McCulloch,
F.~Michel,
R.~M.~Noack,
J.~E.~Paki,
O.~Parcollet,
G.~Paw{\l}owski,
J.~D.~Picon,
T.~Pruschke,
E.~Santos,
G.~Schmid,
U.~Schollw\"ock,
C.~Silva,
F.~St\"ockli,
B.~Surer,
S.~Trebst,
M.~L.~Wall,
and S.~Wessel.


\paragraph{Funding information}
This material is based upon work supported by the National Science Foundation (NSF) under Award No. 2550169 and Award No. 2345575.
Prior support came from ETH Zurich and the Simons Foundation. 
Libraries and application codes integrated in ALPS have been developed over many years with support from
numerous agencies.

\paragraph{Use of generative artificial intelligence:} The authors used AI tools for limited assistance with code completion, testing, verification, and documentation formatting. All generated suggestions were reviewed, modified where appropriate, and independently validated by the authors. The authors take full responsibility for the manuscript and the accompanying codebase.

\bibliography{references}

\end{document}